\documentclass[11pt]{article}
\usepackage{amsmath,amsfonts}
\usepackage[normalem]{ulem}
\usepackage{tikz}
\usetikzlibrary{trees,er,snakes,shapes,mindmap}
\def \beq  {\begin{equation}}
\def \eeq  {\end{equation}}
\def \beqar {\begin{eqnarray}}
\def \eeqar {\end{eqnarray}}
\def\sqr#1#2{{\vcenter{\vbox{\hrule height.#2pt
\hbox{\vrule width.#2pt height#1pt \kern#1pt
\vrule width.#2pt}\hrule height.#2pt}}}}

\def\la {{\langle}}
\def\ra {{\rangle}}
\def\vx {{\vec x}}

\def\vk {{\vec k}}
\def\vf {{\varphi}}

\def\Tr {{\rm Tr}}

\def\vk {\vec{k}}
\def\vx {{\vec x}}
\def\vz {\vec{z}}

\def\vw {\vec{w}}

\def\del {\partial}
\def\bdel{\bar{\partial}}

\def\d {\delta}

\def\bz {{\bar{z}}}

\def\E {{\cal E}}

\def\vf {{\varphi}}

\def\half{\textstyle{1\over 2}}

\begin{document}
\fontfamily{bch}\fontsize{11pt}{15pt}\selectfont
\def \CMP {{Commun. Math. Phys.}}
\def \PRL {{Phys. Rev. Lett.}}
\def \PL {{Phys. Lett.}}
\def \NPBProc {{Nucl. Phys. B (Proc. Suppl.)}}
\def \NP {{Nucl. Phys.}}
\def \RMP {{Rev. Mod. Phys.}}
\def \JGP {{J. Geom. Phys.}}
\def \CQG {{Class. Quant. Grav.}}
\def \MPL {{Mod. Phys. Lett.}}
\def \IJMP {{ Int. J. Mod. Phys.}}
\def \JHEP {{JHEP}}
\def \PR {{Phys. Rev.}}
\def \JMP {{J. Math. Phys.}}
\def \GRG{{Gen. Rel. Grav.}}
\begin{titlepage}
\null\vspace{-62pt} \pagestyle{empty}
\begin{center}
\rightline{CCNY-HEP-18/4}
\rightline{August 2018}
\vspace{1truein} {\Large\bfseries
Casimir Effect in (2+1)-dimensional Yang-Mills Theory as a}\\
\vskip .15in
{\Large\bfseries Probe of the Magnetic Mass}\\
\vskip .1in
{\Large\bfseries ~}\\
{\large\sc Dimitra Karabali$^a$} and
 {\large\sc V.P. Nair$^b$}\\
\vskip .2in
{\itshape $^a$Department of Physics and Astronomy\\
Lehman College of the CUNY\\
Bronx, NY 10468}\\
\vskip .1in
{\itshape $^b$Physics Department\\
City College of the CUNY\\
New York, NY 10031}\\
\vskip .1in
\begin{tabular}{r l}
E-mail:&{\fontfamily{cmtt}\fontsize{11pt}{15pt}\selectfont dimitra.karabali@lehman.cuny.edu}\\
&{\fontfamily{cmtt}\fontsize{11pt}{15pt}\selectfont vpnair@ccny.cuny.edu}
\end{tabular}

\fontfamily{cmr}\fontsize{11pt}{15pt}\selectfont
\vspace{.8in}
\centerline{\large\bf Abstract}
\end{center}
We consider the Casimir effect in a gauge-invariant Hamiltonian formulation of
nonabelian gauge theories in $(2+1)$ dimensions, for an arbitrary gauge group.
We show that the result is in good agreement with recent
lattice simulations. We also argue that the Casimir effect may be viewed as a good
probe of magnetic screening effects in $(3+1)$-dimensional gauge theories at high
temperatures.

\end{titlepage}
\pagestyle{plain} \setcounter{page}{2}
\section{Introduction}

Yang-Mills gauge theories in two spatial dimensions
can be viewed as a guiding model
for the more realistic, but also more 
complicated,
$(3+1)$-dimensional gauge theories. 
The $(2+1)$-dimensional theories have nontrivial dynamical content and 
propagating degrees of freedom making them a better model than
 Yang-Mills theories in 
$(1+1)$ dimensions, yet they are still somewhat more amenable to mathematical 
analysis compared to their
$(3+1)$-dimensional counterparts. 
The Euclidean $3$-dimensional theory, the Wick-rotated version of
the $(2+1)$-dimensional theory, 
can also be of direct relevance to the high temperature 
limit of the $(3+1)$-dimensional theory \cite{high-t}.
In particular, the mass which appears as a propagator mass in 
$(2+1)$ dimensions can be taken as the high temperature value of the magnetic
screening mass.
With these motivations, for many years,
we have been pursuing a Hamiltonian approach to the 
nonpertrubative aspects of Yang-Mills theories in $(2+1)$ dimensions
\cite{{KN1},{KKN1},{nair-trento1}}.
This article will be in the nature of continued work along these lines,
focusing on the Casimir effect
in Yang-Mills theories in
(2+1) dimensions. This was also inspired by the recent lattice simulations of the Casimir effect
for the $SU(2)$ gauge theory reported
in \cite{chernodub}. We will argue that the Casimir effect in the $(2+1)$-dimensional Yang-Mills theory can be viewed as
a probe of the magnetic screening mass in the pure QCD plasma
 in $(3+1)$ dimensions at high temperatures.
This will also furnish a calculation for a general gauge group which can, hopefully,
 be tested in lattice simulations in the near future.

We begin with a brief recapitulation of the salient points of our Hamiltonian analysis.
We considered the $A_0 =0 $ gauge, with the spatial components of the gauge potentials 
parametrized as
\beq
A_z = {\half} (A_1 + i A_2) = - \del M ~M^{-1}, \hskip .2in A_{\bz} = {\half} (A_1 - i A_2 )
= M^{\dagger -1} \bdel M^\dagger
\label{1}
\eeq
Here we use complex coordinates $z = x_1 - i x_2$, $\bz = x_1 + i x_2$ with
$\del= {\half}( \del_1 + i \del_2)$, $\bdel = {\half} (\del_1 - i \del_2)$, and 
$M$ is an element of the complexified group $G^{\mathbb C}$; i.e., it is an $SL(N, {\mathbb{C}})$-matrix
if the gauge transformations take values in $SU(N)$.
Gauge transformations act on $M$ via
$M \rightarrow M^g = g \,M$, where $g$ is an element of the group
$G$, say,  for example, $SU(N)$.
Wave functions are gauge-invariant functionals of $H = M^\dagger M$, with the inner product
given as
\beq
\la 1\vert 2\ra = \int d\mu (H) \exp [2~c_A~S_{wzw}(H)]~ \Psi_1^* \Psi_2
\label{2}
\eeq
Here $S_{wzw}$ is the Wess-Zumino-Witten action given by
\beq
S_{wzw} (H) = {1 \over {2 \pi}} \int \Tr (\partial H ~\bdel
H^{-1}) +{i \over {12 \pi}} \int \epsilon ^{\mu \nu \alpha} \Tr (
H^{-1}
\partial _{\mu} H~ H^{-1}
\partial _{\nu}H ~H^{-1} \partial _{\alpha}H)
\label{3}
\eeq
In equation (\ref{2}), $d\mu (H)$ is the Haar measure for 
$H$ which takes values in $SL(N, {\mathbb{C}})/SU(N)$. Also $c_A$ denotes the value 
of the quadratic Casimir operator for the adjoint representation;
it is equal to $N$ for $SU(N)$.
The Hamiltonian and other observables can be expressed as
 functions of the current $J$ of the WZW action, namely,
\beq
J = {2 \over e} \, \del H ~ H^{-1}
\label{4}
\eeq
(We have included a prefactor involving the coupling $e$; this is useful for later calculations.)
The explicit formulae worked out in references
\cite{{KN1},{KKN1}, {nair-trento1}} is given as
${\cal H} = {\cal H}_0 + {\cal H}_1$, where
\beqar
{\cal H}_0 &=& m  \int_z J_a (\vz) {\d \over {\d J_a (\vz)}} + {2\over \pi} \int _{z,w} 
 {1\over (z-w)^2} {\d \over {\d J_a (\vw)}} {\d \over {\d
J_a (\vz)}}\nonumber\\
&&\hskip .4in + {1\over 2} \int_x :\bdel J^a(x) ~\bdel J^a(x):\label{5}\\
{\cal H}_1& =&  i ~{e}~ f_{abc} \int_{z,w}  {J^c(\vw) \over \pi (z-w)}~ {\d \over {\d J_a (\vw)}} {\d \over {\d
J_b (\vz)}} \nonumber
\eeqar
where $m = e^2 c_A /2\pi$. Regularization issues have been discussed in some detail in the
cited references.

The basic strategy we used was to solve the Schr\"odinger equation keeping all terms in ${\cal H}_0$
at the lowest  order, treating ${\cal H}_1$ as a perturbation. 
In ordinary perturbation theory (carried out using our Hamiltonian formulation), 
one would expand $H= \exp ( t_a \vf^a )$
in powers of the hermitian field $\vf^a$; in addition, since
$m = e^2 c_A/2\pi$ we would also expand
in powers of $m$. In our case,
we keep the term  involving $m$
even at the lowest order.
 So even if we expand $H$ in terms of $\vf^a$,
 our expansion would correspond to a partially resummed version 
 of what would be normal perturbation expansion.
Formally, we keep $m$ and $e$ as independent parameters in keeping track of different orders
in solving the Schr\"odinger equation, only setting $m = e^2 c_A/2\pi$ at the end.
The lowest order computation of the wave function in this scheme was given in \cite{KKN1} and gave the string tension for a Wilson loop in the representation
$R$ as $\sigma_R = e^4 c_A c_R /4 \pi$, $c_R$ being the quadratic Casimir value
for the representation $R$.
 We have also
considered corrections to this formula, taking the expansion to the next higher order (which still involves an infinity of correction terms) and found that the corrections were small, of the order of 
$-0.03\%$ to $-2.8\%$ \cite{KNY}. The resulting values for the string tension
agree well with the lattice estimates
\cite{{teper},{teper2}}.

Some of the other issues explored within this approach include
string breaking effects \cite{AKN}, effective action and ${\mathbb Z}_N$ vortices
\cite{nair2}, supersymmetric theories \cite{AN2},
and entanglement effects \cite{AKN2}. Glueball masses have been discussed in 
\cite{LMY}.

\section{The Casimir energy for parallel wires}

There is an important feature which emerged from our analysis, which is very useful for the present purpose
\cite{KNrobust}.
We can absorb the factor $e^{2 c_A S_{wzw}}$ in (\ref{2}) into the definition of the wave function by writing $\Psi = e^{- c_A S_{wzw}} \, \Phi$.  The inner product for the $\Phi$'s will involve
just the Haar measure without the $e^{2 c_A S_{wzw}}$ factor.
However, the
Hamiltonian acting on $\Phi$ will now be given by
${\cal H} \rightarrow e^{- c_A S_{wzw}}\, {\cal H}\, e^{- c_A S_{wzw}}$.
We can expand $H$ as  $H = \exp( t_a \vf^a ) \, \approx  1 + t_a \vf^a + \cdots$, with the field
$\vf^a$ being hermitian.
As mentioned earlier, this ``small $\vf$" expansion is suitable for a (resummed) perturbation theory. The Hamiltonian is then
\beq
{\cal H}= {1\over 2}\int \left[ -{\delta^2 \over \delta \phi^2} +\phi (-\nabla^2 +m^2)\phi
+\cdots\right] \label{6}
\eeq
where $\phi _a (\vk) = \sqrt {{c_A k \bar{k} }/ (2 \pi m)}~~ \vf _a (\vk)$. 
This is clearly the Hamiltonian for a field of mass $m$ with the corresponding vacuum wave function 
\beq
\Phi_0 \approx \exp \left[ - {1\over 2} \int \phi^a \sqrt{ m^2 - \nabla^2} ~\phi^a \right]
\label{7}
\eeq
The Hamiltonian (\ref{6}) corresponds to the  action
\beq
S = \int d^3x \, {1\over 2}\left[  {\dot \phi}^a {\dot \phi}^a - {(\nabla \phi^a ) (\nabla \phi^a ) }
- {m^2 \phi^a \phi^a }  \right] + \cdots
\label{8}
\eeq

These results show that the propagator for the gauge-invariant component of the gluon field
is the same as that of a massive scalar field with mass equal to $m = (e^2 c_A/2 \pi )$.
Further, the parametrization (\ref{1}) of the gauge potentials 
becomes, in the small $\vf$-expansion
\beq
A^a_i \approx {1\over 2} \left[ - \del_i \theta^a + \epsilon_{ij} \, \del_j \vf^a + \cdots\right], \hskip .2in
M = \exp\left( - {i\over 2} t_a (\theta^a + i \vf^a )\right)
\label{9}
\eeq
In the case of a perfectly conducting plate, the boundary condition is that the tangential
component of the electric field should be zero. In other words, we need
\beq
\epsilon_{ij} \, n_i F_{0j}^a = 0 ,
\label{10}
\eeq
where $n_i$ is the unit vector normal to the
plate. This is also the condition used in
\cite{chernodub}.
In terms of the parametrization
in (\ref{9}), focusing just on the gauge-invariant part $\vf^a$, this means that we need
\beq
n _i \epsilon_{ij} \epsilon_{jk} \del_k {\dot \vf^a} = 
- n_i \del_i {\dot \vf^a} = 0
\label{11}
\eeq 
Since the time-derivative does not affect the spatial boundary conditions, 
this is equivalent to imposing Neumann boundary conditions on the
scalar field $\vf^a$ or, equivalently, on $\phi^a$.
{\it The end result is that, within this approximation of keeping $m$, but expanding the field
$H$ to the lowest order in $\vf^a$, the Casimir energy will be given by that of a
massive scalar field with Neumann boundary conditions on the plates.}

We now consider the standard arrangement of two parallel plates (or rather wires since we are
in two spatial dimensions) which are of infinite extent in the $x_2$-direction and are normal to the
$x_1$-direction. The wires are separated by a distance $R$. We take the range of $x_2$
to be $L$, with $L \rightarrow \infty$ eventually.
The fields in the region between the wires have the mode expansion
\beq
\phi^a = \int {d k\over 2\pi}  \sum_{n=0}^\infty C^a_{n, k} \, \sqrt{2 \over R}
\cos\left( {n \pi x_1 \over R}\right) \, e^{ik x_2}
\label{12}
\eeq
This is consistent with the Neumann boundary conditions.
We note that the Casimir energy of massive scalar fields for the parallel plate geometry
with Dirichlet boundary conditions is known \cite{mass-scalar}. The result for Neumann
conditions is essentially the same. Here we reproduce the result and express it in a form more suitable for comparison with lattice estimates.
With the mode expansion (\ref{12}), the action
(\ref{8}) becomes
\beq
S = \int {d k \over 2 \pi} \sum_n {1\over 2} \left[
{\dot C^a_{n,k} } {\dot C^a_{n,k} } - \Omega_{n,k}^2 \, {C^a_{n,k} }{C^a_{n,k} }
\right] + \cdots
\label{13}
\eeq
where $\Omega^2_{n,k} = k^2 + (n \pi /R)^2 + m^2$. The diagonalization of the
Hamiltonian is trivial, yielding the unrenormalized zero-point energy
\beqar
\E &=& {L \over 2} {\rm dim}G \, \int {d k \over 2 \pi} \sum_n \sqrt {(n \pi /R)^2 + k^2 + m^2}
\nonumber\\
&=& { L \over {2\, \Gamma (-{\half})}}\, {\rm dim}G \, \int {d k \over 2 \pi} \int_0^{\infty} { ds \over s^{3/2}} e^{-s(k^2 + m^2)} \sum_{n=0}^{\infty} e^{-s (n \pi /R)^2} 
\label{14}
\eeqar
Using the Poisson summation formula we get
\beq
\E = { L \over {2\, \Gamma (-{\half})}}\, {\rm dim}G \, \int {d k \over 2 \pi} \int_0^{\infty} { ds \over s^{3/2}} e^{-s(k^2 + m^2)} { 1 \over 2}  \left[ 1 + { R \over {\sqrt{\pi s}}} + 2 \sum_{n=1}^{\infty} {R \over {\sqrt{\pi s}}} e^{-n^2 R^2 /s} \right]
\label{15}
\eeq
The first two terms in this expression are divergent and they have to be subtracted. The first term is independent of the distance $R$ between the wires, corresponds to a self-energy contribution, and gets subtracted when we consider the energy shift 
$\E (R) - \E (R \rightarrow \infty )$, which is the relevant renormalized 
quantity of interest. The second term is proportional to the spatial volume $RL$ and is part of a uniform spatial density of vacuum energy which must also be subtracted out in the renormalized expression for the Casimir energy. 
The final renormalized expression is thus
\beq
\E = { L \over {2 \Gamma (-{\half})}}\, {\rm dim}G \, \int {d k \over 2 \pi} \int_0^{\infty} { ds \over s^{3/2}} e^{-s(k^2 + m^2)}  \sum_{n=1}^{\infty} {R \over {\sqrt{\pi s}}} e^{-n^2 R^2 /s}
\label{15a}
\eeq
Doing the $k$-integration and using the variable transformation $s= (na/m ) e^{\theta}$, we find
\beqar
\E &=& - {L R \over 4} \, {\rm dim}G \, \left({ m \over {\pi R}}\right) ^{3/2} \sum_{n=1}^{\infty} \int { d\theta \over {n^{3/2}}}~\cosh (3 \theta /2)\,  e^{-2nmR\cosh\theta} \nonumber\\
&=& - {L R \over 4} {\rm dim}G\,\left({ m \over {\pi R}}\right) ^{3/2} \sum_{n=1}^{\infty} { K_{3/2} (2nmR) \over n^{3/2}} \label{16}
\eeqar
Using the following expression for modified Bessel function $K_{3/2}$,
\beq
K_{3/2}(z) = \sqrt{\pi \over 2z} e^{-z} \left(1 + {1 \over z} \right)
\eeq 
we can rewrite the Casimir energy as 
\beq
\E = - {\rm dim}G\, {L \over 16\pi R^2}  
\left[  2m R ~ {\rm Li}_2 (e^{- 2 m R}) + {\rm Li}_3(e^{- 2 m R} )\right]
\label{17}
\eeq
where ${\rm Li}_s (w)$ is the polylogarithm function
\beq
{\rm Li}_s (w) = \sum_{1}^\infty  {w^n \over n^s}
\label{18}
\eeq
We may note that, in the $m \rightarrow 0$ limit, the expression (\ref{17})
agrees with the well known result for a massless scalar in $(2+1)$ dimensions,
\beq
\E_{m=0} = - {{ L ~\zeta (3)} \over {16 \pi R^2}}
\label{18a}
\eeq
There are other equivalent ways to arrive at result (\ref{17}). Using
\beq
 \Omega_{n,k} = \left( {\del^2 \over \del x_0^2 }\right)
\int {dk_0 \over \pi} { e^{ik_0 x_0} \over k_0^2 + \Omega_{n,k}^2}  \biggr]_{x_0 =0}
\label{18b}
\eeq
we can carry out the summation over $n$ (in (\ref{14})) to obtain
\beqar
\E &=& - L R^2 \, {\rm dim}G \, \int {d^2 k \over (2\pi)^2} {k_0^2 \over \omega} {1\over e^{2 \omega} -1}
\nonumber\\
&=& - {L R \over 4\pi} {\rm dim}G\,\int_0^\infty dp \, {p^3 \over \sqrt{p^2 + m^2} }\,
{1\over e^{ 2 R \sqrt{p^2 + m^2} } - 1}
\label{18c}
\eeqar
where $\omega^2  = R^2 (k_0^2 +k^2 + m^2)$ and
 in the second line we used polar coordinates in the $(k, k_0)$-plane and integrated
over the angle and $p = \sqrt{k_0^2 + k^2}$. A further substitution
$p R = \sinh q$, and $z = \cosh q$, reduces this to
\beq
\E = - {\rm dim}G\, {L \over 4\pi R^2} (m R)^3 \int_1^\infty d z  { (z^2 -1) \over e^{2 m R  z} - 1}
\label{18d}
\eeq
Expansion of the integrand in powers of $e^{- 2 m R z}$ gives the result
(\ref{17}) in terms of the polylogarithms.

It is useful to write the energy (\ref{17}), for our case, in terms of the string tension 
corresponding to the fundamental representation.
This has been calculated in \cite{{KKN1}}. Ignoring the small corrections
discussed in \cite{KNY}, this is given by
\beq
\sigma_F = e^4\, {c_A c_F \over 4 \pi}
\label{19}
\eeq
We may thus write $m R = \sqrt{c_A /\pi  c_F }\,\, x$, where $x = R\sqrt{\sigma_F} $.
The Casimir energy is thus given by
\beq
{\E \over L \sigma_F } = - {{\rm dim}G \over 16\pi} \left[ {2\sqrt{c_A/\pi c_F} \over x}\,
{\rm Li}_2 \left( e^{- 2\sqrt{c_A/\pi c_F} \,x} \right) + { 1 \over x^2} {\rm Li}_3 \left( e^{- 2\sqrt{c_A/\pi c_F}\, x} \right)
\right]
\label{20}
\eeq
This is the main result of this paper. It holds for an arbitrary compact group;
for the case of $SU(N)$, we have $c_A = N$, $c_F = (N^2 -1) / (2 N)$.
There will be corrections to this formula
due to the fact that we have neglected interactions involving cubic and higher
powers of $\vf^a$ and due to the corrections to the string tension in the expression
for
$m$ in terms of $\sigma_F$. Nevertheless, the fact that string tension given in
(\ref{19}) to
the lowest order in our expansion scheme is in good agreement with lattice 
calculations \cite{teper} suggests that
the formula (\ref{20}) can be a good estimate of the Casimir energy.

\begin{figure}[!b]
\begin{center}
\begin{tikzpicture}[scale = 1.7,domain=0:7]
\pgftext{
\scalebox{.5}{\includegraphics{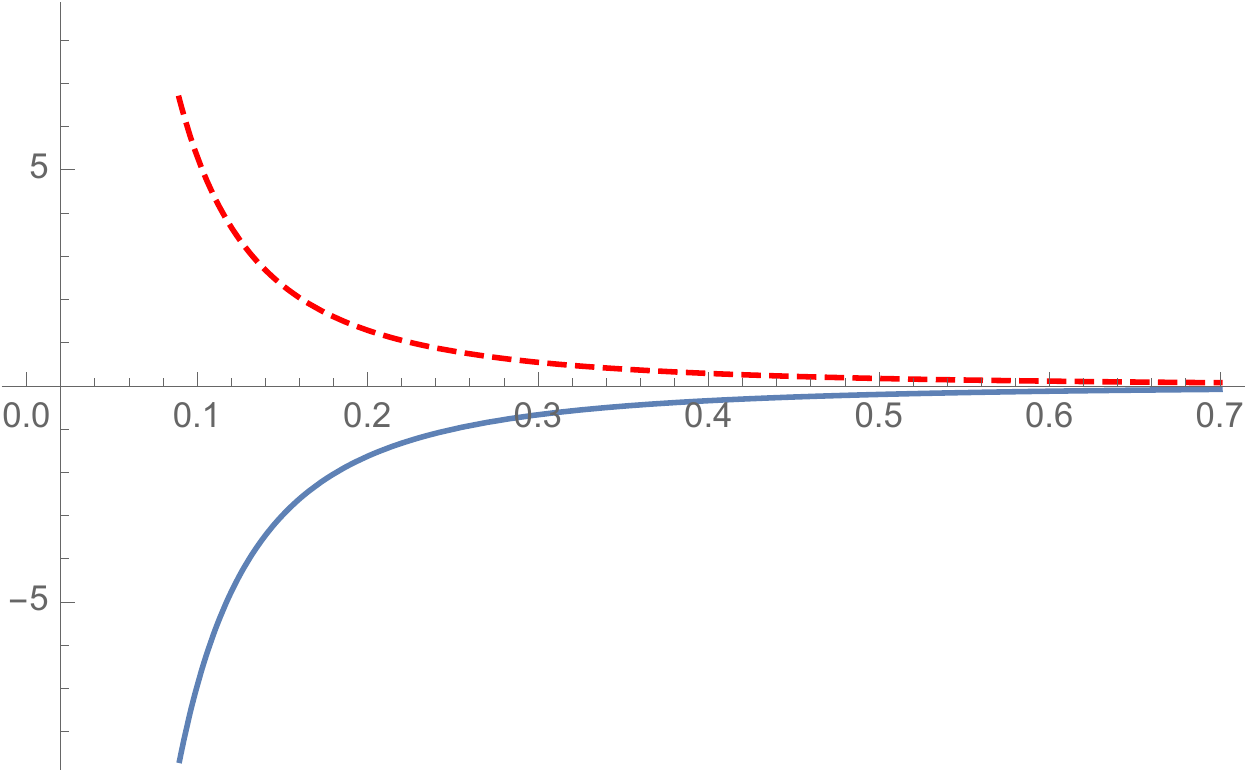}}
}
\draw(-1.7,-1.3)node{${\E_{NN}\over L\sigma_F}$};
\draw(-1.8,1.3)node{${\E_{DN}\over L\sigma_F}$};
\draw(1.5,-.4)node{$x$};
\end{tikzpicture}
\caption{Comparison of $\E_{NN}= \E_{DD}$ from (\ref{20}) (solid line) and $\E_{DN}$ from
(\ref{20b}) (dashed line). }
\label{Graph1}
\end{center}
\end{figure}

We have the Neumann boundary condition on the field $\phi^a$ 
for perfectly conducting wires, as mentioned before. But if we choose different
boundary conditions, the result can be different. Formula
(\ref{20}) holds for the field obeying Neumann conditions at both wires or
Dirichlet conditions at both wires. The Dirichlet condition is equivalent to the magnetic
field $B$ (which is $-\nabla^2 \vf^a$ in our approximation)
vanishing at the wire.
If we consider the Neumann condition at one wire and the Dirichlet condition at the other,
the modes involved are of the form $\sin \left( (n+{\half})\pi x_1/R\right)$.
The Casimir energy is now given by
\beq
\E_{D N} =  {L \over 2} {\rm dim}G \, \int {d k \over 2 \pi} \sum_n \sqrt { ((n+{\half}) \pi /R)^2 + k^2 + m^2}
\label{20a}
\eeq
The renormalized finite Casimir energy now works out to be
\beqar
{\E_{D N}\over L \sigma_F } &=& - {\rm dim}G\, {L \over 16\pi R^2}  
\left[  2m R ~ {\rm Li}_2 (- e^{- 2 m R}) + {\rm Li}_3( - e^{- 2 m R} )\right]\nonumber\\
&=& - {{\rm dim}G \over 16\pi} \left[ {2\sqrt{c_A/\pi c_F} \over x}\,
{\rm Li}_2 \left( -e^{- 2\sqrt{c_A/\pi c_F} \,x} \right) + { 1 \over x^2} {\rm Li}_3 \left( -e^{- 2\sqrt{c_A/\pi c_F}\, x} \right)
\right]
\label{20b}
\eeqar
Notice that, as expected, this corresponds to
a repulsive force because the arguments of the polylogarithms have changed sign.
The two energies $\E$ from (\ref{20}) (same as $\E_{NN} = \E_{DD}$)
and $\E_{DN}$ from (\ref{20b})
are shown in Fig.\,\ref{Graph1} for the case of $SU(2)$.

\section{Lattice estimates and the magnetic mass}

The Casimir energy for the parallel wire geometry was recently evaluated 
for the $SU(2)$ gauge theory by lattice simulation in \cite{chernodub},
with the boundary condition of the tangential component of the electric field vanishing
at the wires. 
(This would be the Neumann-Neumann case for the field $\phi^a$ in our parametrization of $A_i^a$.)
Essentially, the expectation value of the energy density was calculated, 
with a suitable renormalization. The result was fitted to the form
\beq
{\E \over L \sigma_F} = - {\rm dim}G\, { \zeta (3)\over 16 \pi}\, x^{-\nu} \,
e^{- M_{Cas} x / \sqrt{\sigma_F}}
\label{21}
\eeq
The authors find that the best fit values of the parameters are
$\nu = 2.05$ and $M_{Cas} = 1.38\, \sqrt{\sigma_F}$.
The authors also commented on the fact that $M_{Cas}$ is significantly smaller
than the smallest value for glueball mass, which is approximately
$4.7\, \sqrt{\sigma_F}$. The smallness of the exponent is not a surprise from our 
point of view, since the coefficient of $x$ in the exponential 
in (\ref{20}) is $2\,\sqrt{c_A/\pi c_F} = 2\, \sqrt{8/3 \pi} \sim 1.84$ for the case of $SU(2)$. 
This is also, as expected, significantly smaller than what is given by the glueball mass.
While the numerical value
differs from the value for $M_{Cas}/\sqrt{\sigma_F}$ in \cite{chernodub},
 it should be noted that the form
of the function is different as well. The motivation to use (\ref{21}) as a fitting function
for the Casimir energy was that it reduced to the massless formula correctly, 
upon setting $M_{Cas} = 0$ and $\nu =2$. So it may be viewed as a two-parameter
extension of the formula for the massless case.
Our formula (\ref{20}) also correctly reduces to the massless limit, and so one may contemplate
a modification of (\ref{20}) with additional parameters to be used as a fitting function. One could consider, for example, changing the prefactor in (\ref{20}); there is a reasonable argument for this.
Notice that the prefactor is a measure
of the number of degrees of freedom, as evidenced by the
${\rm dim}G$ factor. Lattice simulations of QCD shows that
the number of degrees of freedom do not quite reach a value corresponding to
a gas of free gluons even at very high temperatures, where we expect a
deconfined gluon plasma.
(This has been known for a while; a recent review which 
gives updated results is \cite{pasechnik}; in particular,
see figure 4 of this reference.) In our calculation presumably such an effect can arise from higher-order terms in
$\vf^a$  which have been neglected.

\begin{figure}[!t]
\begin{center}
\begin{tikzpicture}[scale = 1.4,domain=0:7]
\pgftext{
\scalebox{.5}{\includegraphics{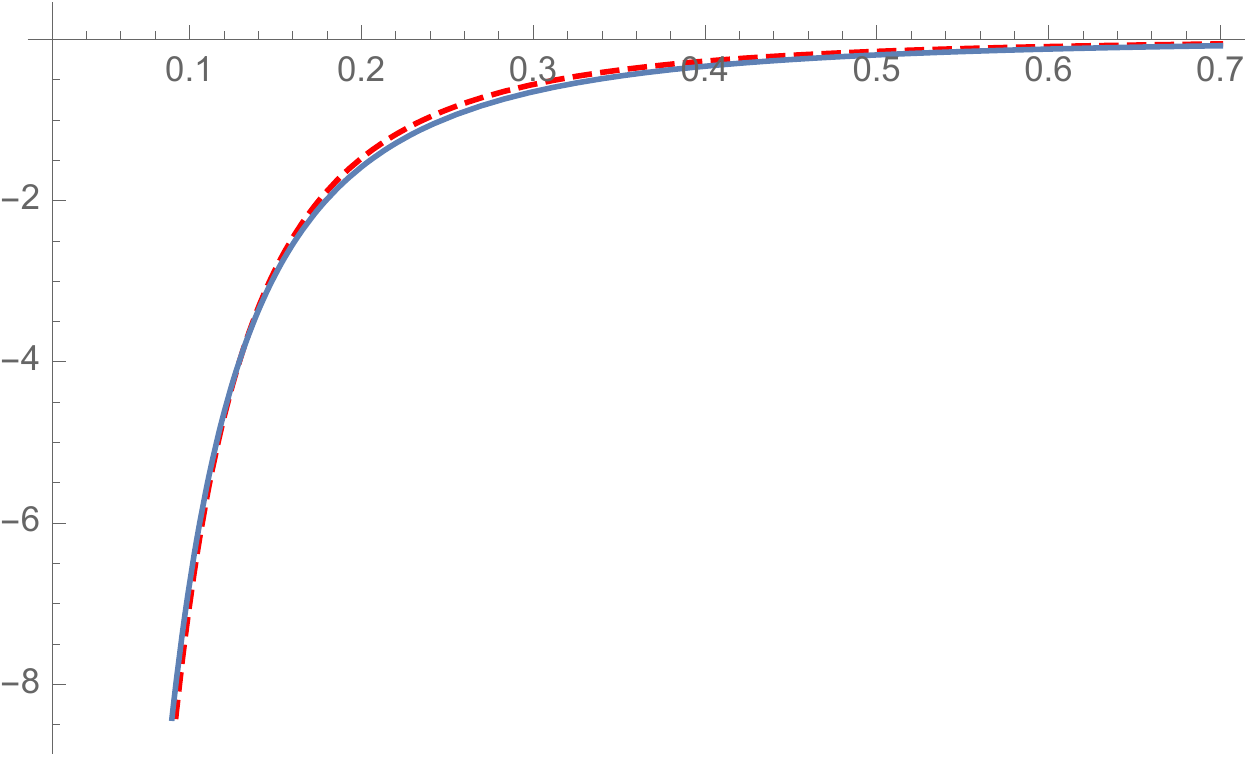}}
}
\draw(-3.5,.5)node{${\E\over L\sigma_F}$};
\draw(1,2)node{$x $};
\end{tikzpicture}
\caption{Comparison of (\ref{21}) (dashed line) and (\ref{22}) (solid line). }
\label{Graph2}
\end{center}
\end{figure}

Another observation is that our calculation based on the previous Hamiltonian analysis 
shows, as explained in more detail later, that there are strong
theoretical reasons why the expression for the Casimir energy should involve powers of $e^{-2mR}$ where $m$ is the
mass as it appears in the propagator for the gauge-invariant part of the gauge potential.
So here we will keep the value of $m$
as the one given by the Hamiltonian analysis, not treated as a parameter 
to be obtained from fitting.
However, based on what was said earlier, we shall use a prefactor and try to
fit the lattice calculation to the formula
\beq
{\E \over L \sigma_F } = - A\,{{\rm dim}G \over 16\pi} \left[ { 1.84 \over x}\,
{\rm Li}_2 \left( e^{- 1.84\,x} \right) + {1 \over x^2} {\rm Li}_3 \left( e^{- 1.84\, x} \right)
\right]
\label{22}
\eeq
where $A$ is to be treated as fitting parameter and we have also put in the values of 
$c_A$, $c_F$ for $SU(2)$. A comparison of (\ref{21}) and (\ref{22}),
with the best fit values $\nu = 2.05$, $M_{Cas} = 1.38 \, \sqrt{\sigma_F}$
for (\ref{21}) and $A= 0.9715$ for (\ref{22}), is shown in Fig.\,\ref{Graph2}.
The range of $x$ is taken to be $0.1$ to $0.7$ as done in
\cite{chernodub}.
The graph shows clearly
that our formula does capture the lattice calculation
of the Casimir energy with good quantitative accuracy.

It is worth emphasizing the significance of the gauge-invariant
Hamiltonian analysis  we have used here. {\it A priori,} it is not clear that the Casimir 
effect for the nonabelian theory
can be reduced to that of a massive scalar field. Our approach shows that this 
can indeed be done. Secondly, we get a specific value for the propagator mass
$m$, namely, $e^2 c_A /(2\pi)$, as well as its relation to the
string tension, since we also have an independent prediction for $\sigma_F$.
 Taking this value, without determining it via fitting to
lattice data, we get good agreement. We have used an overall coefficient $A$ as a
parameter determined by fitting. But the best fit value is $0.9715$, so that in retrospect, 
we see that even if we took $A$ to be $1$, as it is in our lowest order calculation, the
agreement is still within a few percent.

The good agreement between the lattice results in \cite{chernodub} and our analytical expression (\ref{20}) for the Casimir energy provides yet another strong indication ( in addition to the string tension agreement \cite{{teper}, {teper2}}) that our Hamiltonian analysis, in particular the quadratic approximation, provides a good effective description for (2+1)-dimensional Yang-Mills theory.

The mapping of the Casimir energy to that of a massive scalar field has been
discussed in \cite{chernodub2}\footnote{We thank M. Chernodub for bringing these
papers to our attention.}
for compact Abelian electrodynamics in three dimensions, where the monopoles 
are responsible for the mass generation.
Our approach justifies a mapping to the massive scalar for the nonabelian Yang-Mills theory,
and also yields predictions for
$m$ and $\sigma_F$.

A few more comments on the formula for the Casimir energy
are in order at this point.
First of all, there is an intuitive reasoning for
the exponential dependence on $x$
which is as follows.
The expectation value of the energy
involves the propagator since
\beq
\la \E \ra \sim \int  {\del \over \del x_0} {\del \over \del x'_0} \la A_i^a(\vx, x_0)
A_i^a (\vx, x'_0)\ra \Bigr]_{x_0 = x'_0} \, +\cdots
\eeq
The propagator $\la A_i^a(\vx, x_0)
A_i^a (\vx, x'_0)\ra$ may be viewed in terms of paths from $\vx$ to one of the wires, 
from there to the other wire, and then back to $\vx$.
This involves a distance of $2R$, and with a propagator mass
of $m$, we expect a factor $e^{- 2 m R}$. This should hold
for all boundary conditions for large $R$.
Multiple transits can lead to the formula
with the summation as in the polylogarithm. This argument, as well as our
explicit calculation, makes it clear that the mass in the propagator is what controls
the exponential factor. Of course the precise functional dependence of the Casimir energy on $e^{-2mR}$ depends on the boundary conditions, as displayed for example in equations (\ref{20}) and (\ref{20b}).

Secondly, we note that the
 propagator mass is also related to the magnetic screening mass
 in one higher dimension.
If we consider the $(3+1)$-dimensional Yang-Mills theory at very
high temperatures, in the imaginary time formalism,
all modes except for the lowest Matsubara frequency decouple
and the theory is expected to reduce to a three-dimensional one with
$e^2 = g^2 T$, where $g$ is the $4$-d coupling and $T$ is the temperature.
The mass which appears in the propagator of the 
Euclidean $3$-dimensional theory
then serves as the magnetic screening
mass of the high temperature $(3+1)$-dimensional
theory. For this reason, we often refer to the propagator mass
in our calculation as the magnetic mass.

As for the values of the magnetic mass,
our Hamiltonian calculation gives $m = e^2 c_A / 2\pi = e^2/\pi \approx 0.32\,e^2$, 
for $SU(2)$.
There have been many 
other ways of estimating the magnetic mass.
These include various resummation and gap equation approaches
\cite{{mass},{AN},{BP}, {JP2}}, lattice analyses in different gauges
\cite{karsch1} and a method of identifying
the magnetic mass as a common divisor for glueball masses
\cite{owe2}. The values obtained are
close to what we find, generally in the range $0.28$ to $0.38$
for $m/e^2$, with the lattice values being somewhat higher,
close to $0.5$. There is general consistency 
among the values, none of them is close to the glueball masses.
(All numerical values with a short discussion are given
in \cite{nair-trento1}.)
Since there is some variation, one could also envisage
the mass as a fitting parameter, although our experience with
the string tension suggests that
the Hamiltonian approach should be closest to lattice simulations.

Finally, this discussion shows that the Casimir effect in the $(2+1)$-dimensional
theory is a good probe of the magnetic screening mass
for the $(3+1)$-dimensional theory. 
Lattice simulations
for groups other than $SU(2)$ and comparison with our formula
(\ref{20}), ( or (\ref{20b}) for mixed boundary conditions) 
with perhaps a prefactor $A$ to be determined, will be
worthwhile, in terms of providing
greater insights into this issue.

\bigskip

We thank Hans Hansson for bringing the paper \cite{chernodub} to our attention.
We also thank Maxim Chernodub for discussions.
This research was supported in part by the U.S.\ National Science
Foundation grant PHY-1519449
and by PSC-CUNY awards.


\end{document}